\documentclass[aps,prb,preprint,groupedaddress,epsfig,showpacs]{revtex4}
\usepackage{graphicx}

\begin{document}

\title{Electronic structure of few-electron concentric double quantum rings}

\author{J.I. Climente}
\affiliation{S3 CNR-INFM, Via Campi 213/A, 41100 Modena, Italy}
\affiliation{Departament de Ci\`encies Experimentals, Universitat Jaume I, 
Box 224, E-12080 Castell\'o, Spain}
\author{J. Planelles}
\email{planelle@exp.uji.es}
\affiliation{Departament de Ci\`encies Experimentals, Universitat Jaume I, 
Box 224, E-12080 Castell\'o, Spain}
\author{M. Barranco, F. Malet, M. Pi}
\affiliation{Departament ECM, Facultat de F\'{\i}sica, Universitat de
Barcelona, E-08028 Barcelona, Spain}

\begin{abstract}
The ground state structure of few-electron concentric double quantum 
rings is investigated within the local spin density approximation. 
Signatures of inter-ring coupling in the addition energy spectrum are 
identified and discussed.
We show that the electronic configurations in these structures can be
greatly modulated by the inter-ring distance: At short and long distances the low-lying
electron states localize in the inner and outer rings, respectively, and
the energy structure is essentially that of an isolated single quantum ring. However, at 
intermediate distances the electron states localized in the inner and the outer ring become 
quasi-degenerate and a rather entangled, strongly-correlated system is formed.
\end{abstract}


\maketitle
\section{Introduction}

Recent experimental works have achieved the formation of `artificial diatomic molecules' made
of self-assembled, strain-free, concentrically-coupled GaAs/AlGaAs quantum rings.\cite{ManoNL}
Understanding the electronic properties of these nanostructures is essential for their eventual 
application in practical devices.
This has sparked a number of early experimental and theoretical studies on their optoelectronic properties.
Photoluminescence spectra of concentric double quantum rings (DQRs) were measured and interpreted as 
evidence of exciton localization in either the inner or the outer ring.\cite{ManoNL,ManoPRB} 
The single-electron\cite{FusterBJP,PlanellesEPJB} and electron-hole\cite{PlanellesEPJB} energy levels in a 
magnetic field were studied theoretically and it was shown that, even for small inter-ring 
distances, self-assembled DQRs could be approximately described as a sum of two decoupled rings.
Szafran and Peeters also investigated the magnetic response of one- to three-electron energy levels in 
DQR structures.\cite{SzafranPRB} They adopted ring dimensions corresponding to structures
produced
by tip-oxidation technique\cite{FuhrerNAT} and, due to their weaker spatial confinement, inter-ring 
tunneling was enhanced as compared to that of self-assembled DQRs.\cite{PlanellesEPJB}

In this paper we extend the previous works by investigating the ground state structure of 
self-assembled DQRs as a function of the inter-ring distance and the number of confined 
conduction band electrons, $N$. Particular emphasis is placed on the role of Coulomb interactions.
Experimentally, the shell structure of quantum dots and quantum rings is often inferred from their
addition energy spectrum,\cite{TaruchaPRL,LorkePRL} so we will use this magnitude in order to
illustrate the energy structure of the systems under study.
The singular geometry of DQRs is expected to introduce characteristic features in the addition 
energy spectrum as compared to that of other artificial molecules. For example, unlike laterally
coupled quantum dots, DQRs couple concentrically and thus preserve their circular symmetry. 
Differences with respect to vertically-coupled\cite{MaletXXX} quantum rings 
are also expected.  Several physical reasons are responsible for this.
First of all, the DQR molecule is `heteronuclear', the volume of the outer ring usually exceeding
that of the inner ring.\cite{ManoNL}  This makes the maximum charge density of bonding and antibonding states 
localize in different rings, therefore showing distinct energy spacing between consecutive azimuthal 
levels.\cite{PlanellesEPJB}
The localization of the electrons in either ring follows from an intrincate interplay between spatial confinement, 
centrifugal forces (which favor the occupation of the inner ring) and Coulomb forces 
(which tend to favor the occupation of the more voluminous ring, as long as it is not heavily charged).

As we shall see, each of these factors prevails in a given range of inter-ring distances. For small
inter-ring separations, a large number of electrons can be placed in the inner ring because Coulomb 
interactions hardly compensate for the stronger vertical confinement of the outer ring and the centrifugal 
stabilization.
On the contrary, with increasing separation the relative volume of the outer ring grows and at some point the electrons 
move into it rather swiftly. Finally, an intermediate distance is found where the single-electron states
localized in the inner and outer rings are close in energy.  
In this latter case, Coulomb interactions become critical to determine the shell filling and electron 
localization. As a matter of fact, they may lead to interesting spin-dipolar configurations, where the 
inner ring charge density is strongly spin-polarized whereas the outer ring one is not.

\section{Theoretical considerations}

Local Spin Density Functional Theory (LSDFT), which has given satisfactory results in  the study of
related structures,\cite{PiPRB,PiPRL,WensauerPRB} 
is employed in the present work. Details of the method followed are described in Ref.~\onlinecite{PiPRB}.
The confining potential in the $(\rho,z)$ plane (i.e., the cross-section profile), is the same we used 
in a previous study of DQRs,\cite{PlanellesEPJB} namely a superimposition of two Gaussian curves, 
one with constant radius $R_{in}$ for the inner ring, and another with changing radius $R_{out}$
for the outer ring. Thus, the height of the DQR is given by:

\begin{equation}
\label{eqH}
H(\rho)=h_{in} \exp \left[
-\left(\frac{\rho-R_{in}}{\sigma_{in}}\right)^2
\right] +h_{out} \exp \left[-\left(\frac{\rho-R_{out}}{\sigma_{out}}\right)^2\right], \end{equation}

\noindent where $h$ is the maximum height of each ring and $\sigma^2=\Delta\nu_{1/2}^2/\ln 2$, $\nu_{1/2}$ 
being the half-width. Note that the height of the two curves adds up when they overlap, so that the area 
of the DQR cross-section is constant.  The confinig potential is then defined as:

\begin{equation}
\label{eqV}
V(\rho,z)=\left\{ \begin{array}{ll}
0 & {\rm if} \; 0 \leq z \leq H(\rho)\\
V_c & {\rm otherwise,} \end{array} \right.
\end{equation}

\noindent where $V_c$ stands for the heterostructure band-offset.
It is worth stressing that this confinement potential allows us to fit accurately the DQR profile observed by
atomic force microscopy in Ref.~\onlinecite{ManoNL}. It also renders a detailed description of the vertical
confinement, which is essential because of the sensitivity of the energy spectrum to the depth of the valley separating 
the inner and outer rings.\cite{PlanellesEPJB} Moreover, within this model, the evolution from a single to a double
quantum ring as $R_{out}$ increases implies a transfer of volume from the inner to the outer ring, which
mimics the As-flux controlled self-assembly realistically.\cite{ManoNL} 
On the other hand, the three-dimensional Hamiltonian -although axially symmetric- is important not to 
overestimate the role of Coulomb interactions.\cite{RontaniAPL} 

We study a set of DQRs similar to those synthesized in Ref.~\onlinecite{ManoNL}. The inner ring radius
is fixed at $R_{in}=22.5$ nm and the outer one is varied from $R_{out}=22.5$ to $R_{out}=50$ nm.
The inner (outer) ring half-width is $12.5$ ($30$) nm, and both rings have $h_{in/out}=4$ nm height.
We take material parameters corresponding to a GaAs/Al$_{0.3}$Ga$_{0.7}$As system:  the effective mass 
is $m^*=0.067$, the dielectric constant $\epsilon=12.4$, and the band-offset $V_c=262$ meV.

As mentioned above, it has been shown that the single-electron states of self-assembled DQRs can be approximately
described as a simple sum of the states of the inner and outer rings, considered as isolated 
entities.\cite{FusterBJP,PlanellesEPJB}
The same occurs in our present calculations, where most Kohn-Sham orbitals localize to a 
great extent in either the inner or the outer ring.
Therefore, for simplicity throughout this paper we will adopt a fuzzy logic language and 
classify the DQR states into inner and outer ring states, the criterium for such classification being
the charge density maximum localization.

\section{Results and discussion}

\begin{figure}[p]
\centerline{\includegraphics[width=8.5cm,clip]{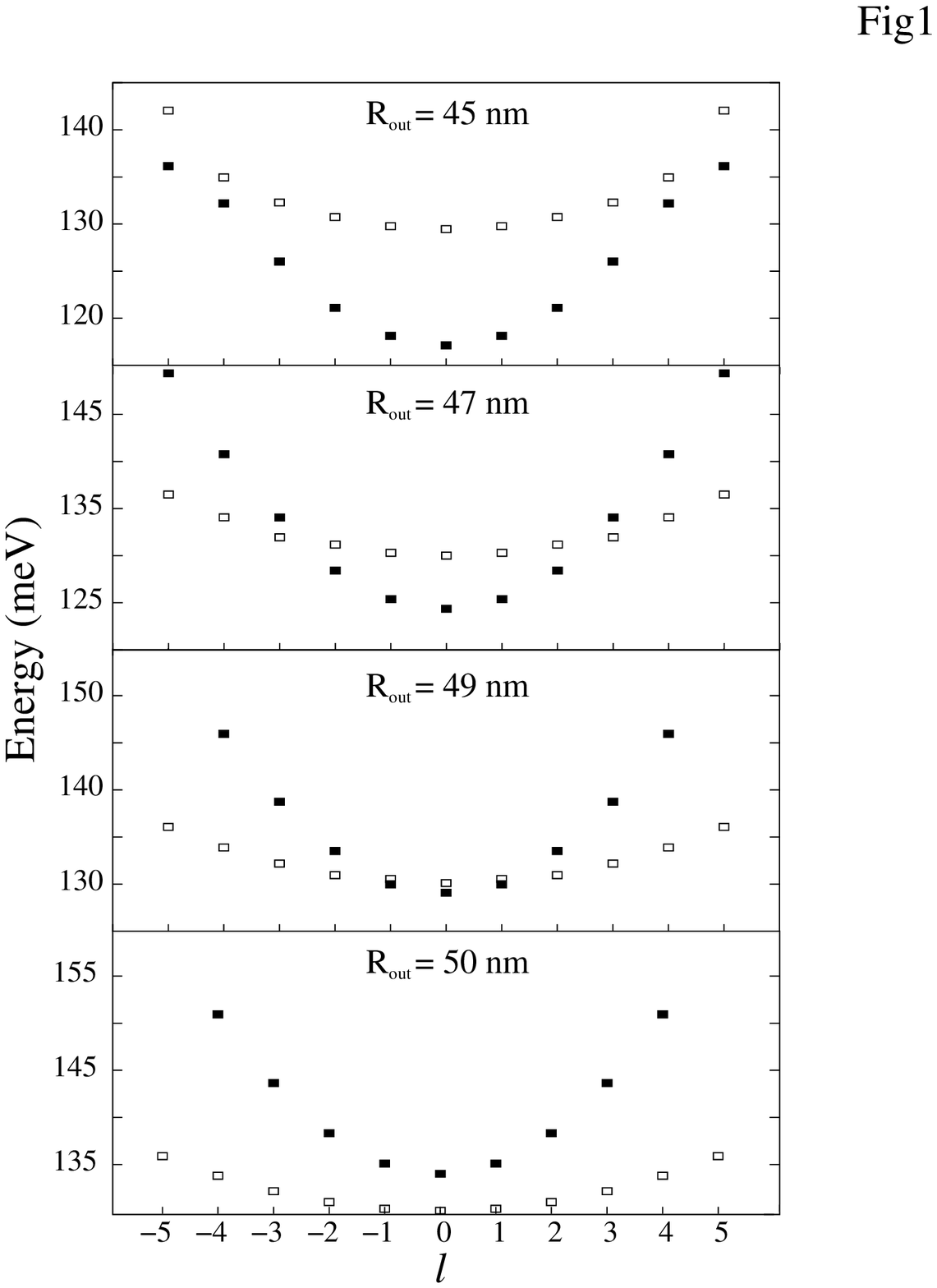} }
\caption{Lowest-lying orbital energy levels vs. angular momentum in DQRs with $R_{in}=22.5$ nm 
and changing $R_{out}$. Coulomb interaction is not included. Solid boxes correspond to states localized
in the inner ring, and open boxes to states localized in the outer ring.}\label{Fig1}
\end{figure}

The electron localization in a DQR is determined by three factors: the centrifugal potential, 
the spatial confinement and the Coulomb repulsion in each ring.
It is convenient for our analysis to disentangle the latter factor by investigating single-particle effects in
the first place. 
Thus, in Figure \ref{Fig1} we show the single-electron orbital energy levels for a number of DQRs 
with increasing $R_{out}$. In the figure, $l$ stands for the (azimuthal) angular momentum and only the two lowest
eigenvalues ($n=0$ and $n=1$) for each $l$ are depicted ($n=2$ states are much higher in energy).
In all cases, $n=0$ and $n=1$ energy levels show clearly different spacing between consecutive azimuthal levels.
This reflects the different mean radii of their charge density, which allows us to distinguish states localizing
in the inner and outer rings (solid and open boxes, respectively).
 For $R_{out}=45$ nm, $n=0$ states localize in the inner ring, whereas $n=1$ ones localize in the outer ring.
As $R_{out}$ increases the relative volume of the inner ring is reduced, and therefore the states localizing in it
are destabilized.  Consequently, with increasing $R_{out}$ the lowest-lying levels of the inner ring first become 
quasi-degenerate with those of the outer ring and finally become more excited 
(i.e., $n=0$ states localize in the outer ring).\\

\begin{figure}[p]
\includegraphics[width=14cm,clip]{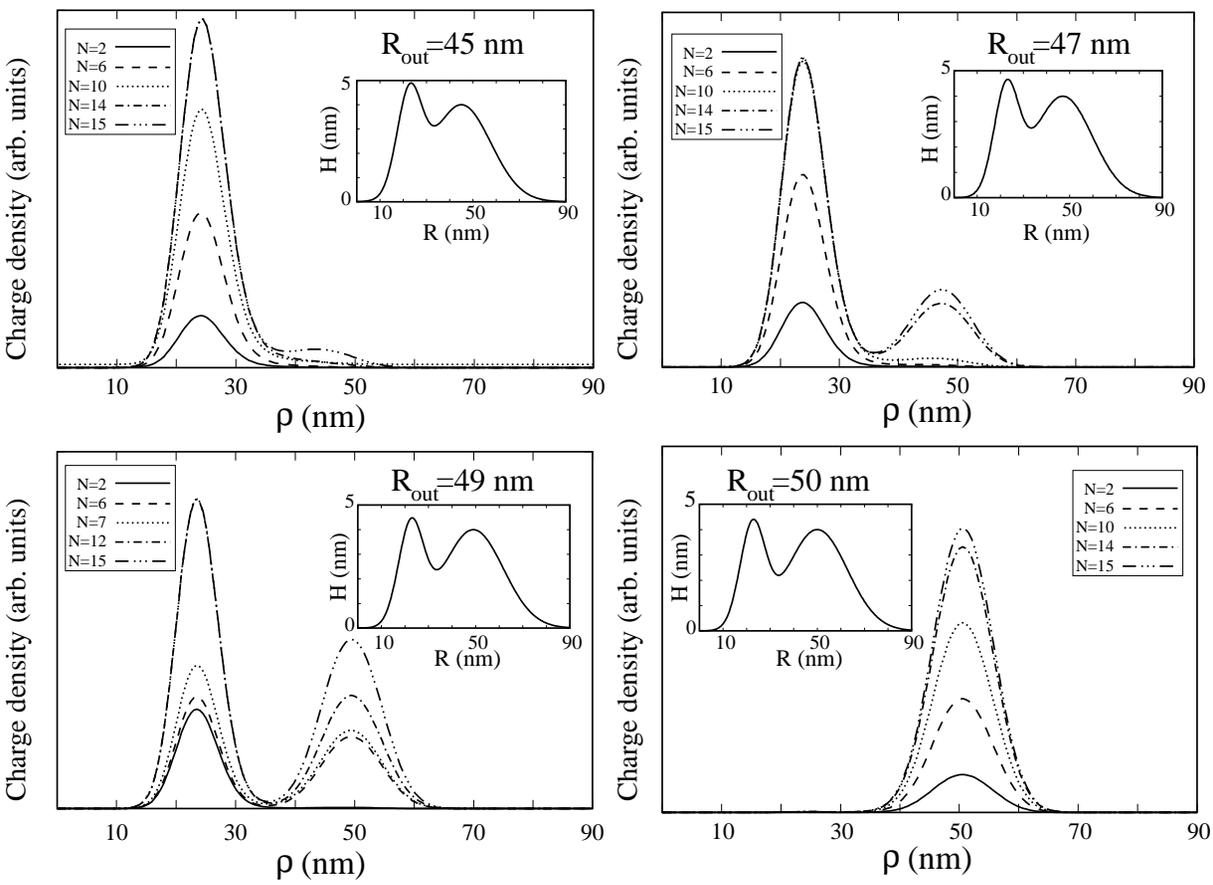}
\caption{Radial charge density distribution of $N$ non-interacting electrons in DQRs with $R_{in}=22.5$ nm and 
changing $R_{out}$. The insets illustrate the DQR cross-section profile.}\label{Fig2}
\end{figure}

Figure \ref{Fig2} illustrates the radial charge density for some $N-$electron ground states in the DQRs of 
Fig.~\ref{Fig1}, which result from an independent-particle filling of the energy levels.
The insets show the corresponding confinement potential profile for each value of $R_{out}$.
It is noted that for $R_{out}=45$ nm, even though the outer ring is clearly formed, almost no leaking of 
the density from the inner to the outer ring is observed. This means that all $n=0$ orbitals are mostly 
localized within the inner ring regardless of their angular momentum. A similar situation occurs 
for $R_{out}=50$ nm, but in this case $n=0$ orbitals are mostly localized in the outer ring. 
Only in the intermediate region, where some $n=1$ states are close in energy to $n=0$ states with $l>0$, 
simultaneous charging of both rings appears. These results indicate that tunneling between the two rings 
is strongly suppressed by the Gaussian-like profile of Ref.~\onlinecite{ManoNL} DQRs cross-section:
since the vertical confinement is much stronger than the lateral one, small differences in the height 
of the inner and outer ring have a dramatic effect on the corresponding energy levels. 
For $R_{out}=45$ nm the outer ring is clearly defined, but its height is lower than that of the inner ring
(see insets of Fig.~\ref{Fig2}), so its energy levels are relatively very excited. On the contrary, 
for $R_{out}=50$, when the height of both rings is already comparable, the outer ring has become much 
wider than the inner ring and therefore its energy levels are more stable. 
It is worth stressing again that these effects cannot be found if the confining potential employed does not 
consider properly the variations of the vertical confinement for each radial position. 
Indeed, we have carried out calculations using a confining potential similar to that of previous works 
for laterally coupled quantum dots\cite{WensauerPRB} and large DQRs\cite{SzafranPRB}, namely
a quantum well in the growth direction and two overlaping parabolae in the radial direction. The results
are then qualitatively different: for all inter-ring distances the $n=0,l=0$ ($n=1,l=0$) states 
localize mainly in the inner (outer) ring, while the $l>0$ states do so in the opposite ring. 
This would suggest that the centrifugal potential and the inter-ring spatial confinement have comparable 
contributions in the Hamiltonian.\cite{SzafranPRB} It does not seem to be the case for self-assembled
DQRs.\cite{FusterBJP,PlanellesEPJB} 

\begin{figure}[p]
\centerline{\includegraphics[width=8.5cm,clip]{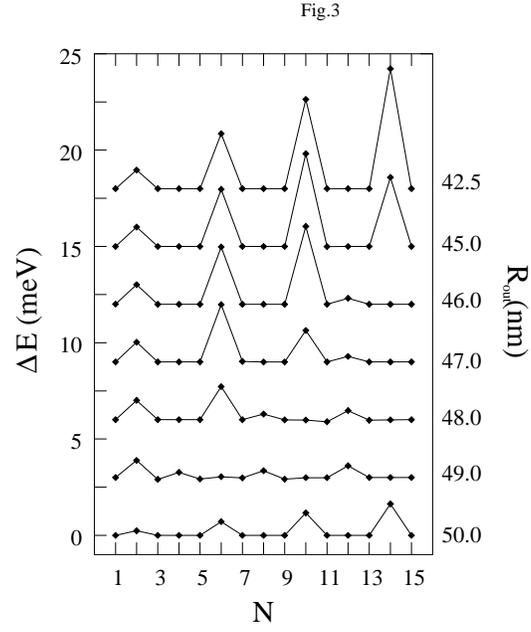} }
\caption{Addition energy spectra versus the number of confined non-interacting electrons 
in DQRs with $R_{in}=22.5$ nm and changing $R_{out}$. The spectra are offset for 
clarity.}\label{Fig3}
\end{figure}

Figure \ref{Fig3} represents the independent particle addition energy spectra
\begin{equation}
A(N)=E(N+1) - 2E(N) + E(N-1) \;\; ,
\label{eq1}
\end{equation}
where $E(N)$ is the energy of the $N$-electron ground state.
For $R_{out}=42.5$ nm the spectrum is essentially that of a single quantum 
ring,\cite{LinPRB,Emperador01} with peaks at
closed-shell configurations $N=2,6,10,14$. Obviously, no peaks are observed at half-shell filling values of
$N$ because we are neglecting Coulomb interactions so far. It can be seen that the height of consecutive maxima
increases with $N$ because the larger $l$ is, the larger the energy spacing $\Delta E(n,l\pm 1)$ becomes 
(see Fig.~\ref{Fig1}).
At $R_{out}=45$ nm a first irregularity is observed: the $N=14$ peak is lower than the $N=10$ one.
This happens because once the $n=0,l=3$ shell is closed by the 14-th electron, the next energy level is not 
$n=0,l=4$ but $n=1,l=0$, which mostly localizes the wave function in the already voluminous external ring.
As $R_{out}$ keeps increasing, the lowest $n=1$ states start catching up with high-angular-momentum $n=0$ states.
This is seen in the spectra as the gradual destruction of the regular single quantum ring pattern, which
is replaced by  lower  peaks at ever smaller values of $N$. The reduced height of the incoming peaks is in part
due to the smaller energy spacing between consecutive azimuthal levels in the outer ring. However, one should also
take into account that the outer ring levels intermix with the inner ring ones, so that consecutive electrons may
fill shells of different rings and, therefore, the height and distribution of the addition energy peaks
is not simply that of the inner ring up to some value of $N$, plus that of the outer ring for larger $N$.
Instead, we generally observe a regular quantum ring spectrum up to the filling of the last-but-one shell 
prior to the $n=1,l=0$ state, and afterwards the peaks become irregular both in height and position.
The most complicated spectrum is found at $R_{out}=49$ nm, when the lowest levels of the inner and outer rings are
quasi-degenerate. A further increase in the inter-ring separation, from $R_{out}=49$ to $R_{out}=50$ nm,
already retrieves the regular spectrum of a single quantum ring. This is because the density of
states in the outer ring is much higher than that of the inner ring. As a consequence, a slight stabilization of the
outer ring levels rapidly leads to a situation where many electrons can be hosted by the outer ring 
before reaching the first inner ring level (see Fig.~\ref{Fig1} for $R_{out}=50$ nm).

We next investigate the influence of Coulomb interactions on DQRs. Since the inner and outer rings have
different volumes, electrons will experience different Coulomb interaction strength in each ring
(we may say that the two rings have different `electroaffinities' to draw a parallel with ordinary molecules).
In general, we expect that Coulomb repulsion pushes electrons towards the larger ring,\cite{SzafranPRB} 
but this trend may be reversed when the larger ring contains too many electrons.

\begin{figure}[p]
\includegraphics[width=14cm,clip]{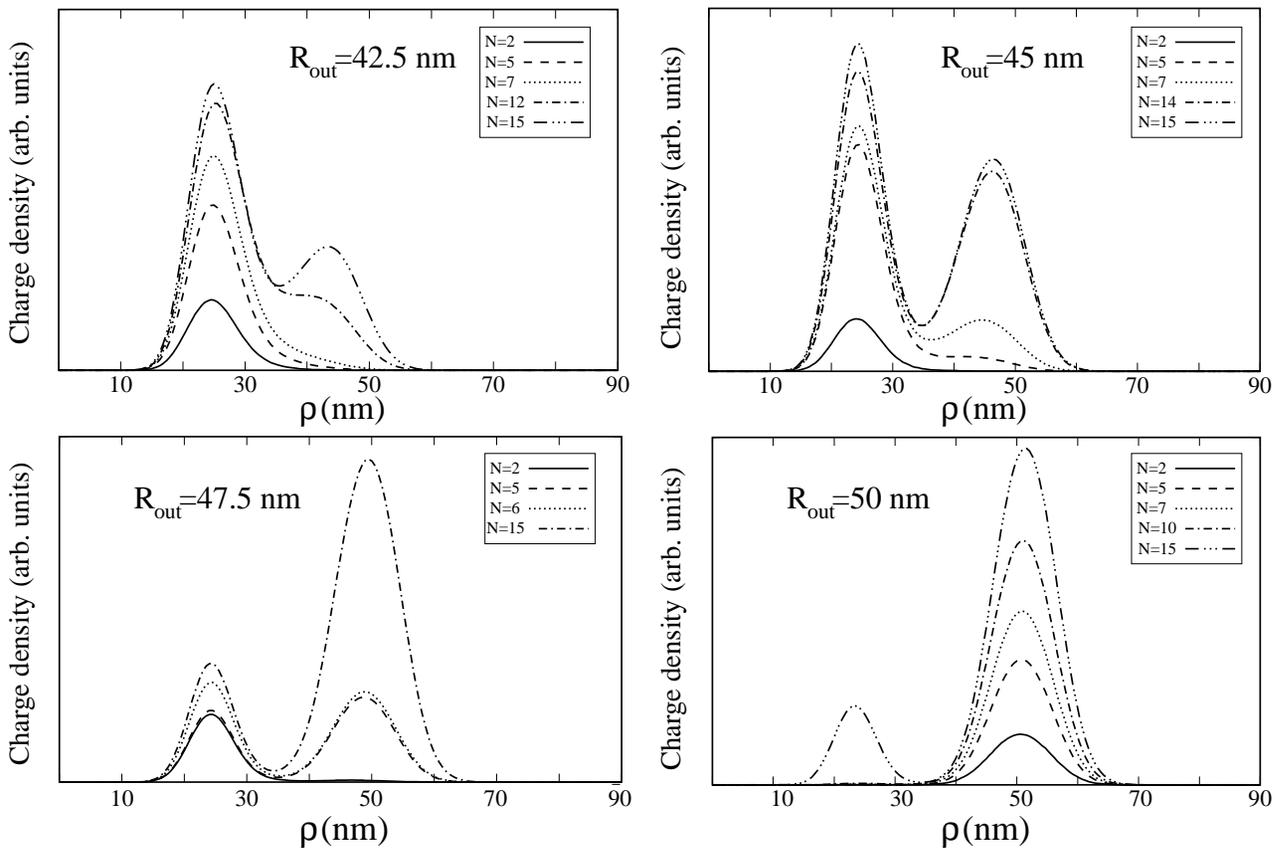}
\caption{Radial charge density distribution of $N$ interacting electrons in DQRs with $R_{in}=22.5$ nm and 
changing $R_{out}$.}\label{Fig4}
\end{figure}

Figure \ref{Fig4} depicts the radial charge densities of a number of $N-$electron ground-states in DQRs
with increasing $R_{out}$, taking direct Coulomb, exchange and correlation contributions into account.  
We observe that charging of  the outer ring starts at smaller values of $R_{out}$ than in the independent 
particle case, because Coulomb interaction helps to compensate for the stronger vertical confinement 
of the outer ring.
Indeed, for $R_{out}=45$ nm, the $N=7$ to $N=14$ electrons already localize in the outer ring, whereas in
the independent particle scheme this only happens from the $N=15$ electron on. If we inspect the
single-particle energy levels of $R_{out}=45$ nm in Fig.~\ref{Fig1} we notice that, for the $N=7$ electron to
fill the lowest $n=1$ state, it skips as much as two empty shells of the inner ring 
($n=0,l=2$ and $n=0,l=3$).  This conspicuous violation of the Aufbau principle is made possible by 
the large difference of Coulomb interaction strength in each ring as compared to the difference in 
kinetic energies.\cite{jp_dixit}
Kinetic terms in the inner ring are smaller than in the outer ring due to the weaker vertical confinement,
but Coulomb repulsion is stronger due to the smaller volume  (see inset in Fig.~\ref{Fig2}).
We also observe that the $N=15$ electron localizes again back in the inner 
ring, because of the accumulated electron charge in the outer ring. 
Likewise, in the $R_{out}=50$ nm picture, despite of the larger volume of the outer ring, Coulomb interactions 
induce the localization of high-$N$ states in the inner ring. 
Nonetheless, the most complicated frame is found at $R_{out}\sim 47.5$ nm, where the Coulomb energy 
stabilization provided by the electron localization in the outer ring is of the same order as the charge
confinement energy stabilization ensuing from localization in the inner ring.
As a result, the charge density localization is extremely sensitive to the number of confined
electrons. Hence, the $N=1-2$ electrons localize in the inner ring, the $N=3-5$ do so in the outer ring, 
the $N=6$ electron is inside again $\ldots$
In other words, around this inter-ring distance the Kohn-Sham orbitals localized in the inner and outer rings
are very close in energy, hence yielding a strongly correlated system and Coulomb-mediated tunneling between 
the two rings becomes very efficient.
It is worth stressing that the entanglement in this structure arises from
`molecular orbitals' with similar energies but very different spatial distributions.
Even though a similar situation may appear in vertically coupled heteronuclear quantum rings\cite{MaletXXX}, 
 it would be difficult to tune (experimentally) the appropiate barrier thickness 
for the dimensions of the constituent rings. In contrast, for DQRs this situation follows 
naturally from the As-flux controlled synthesis described in Ref.~\onlinecite{ManoNL}.

\begin{figure}[p]
\centerline{\includegraphics[width=8.5cm,clip]{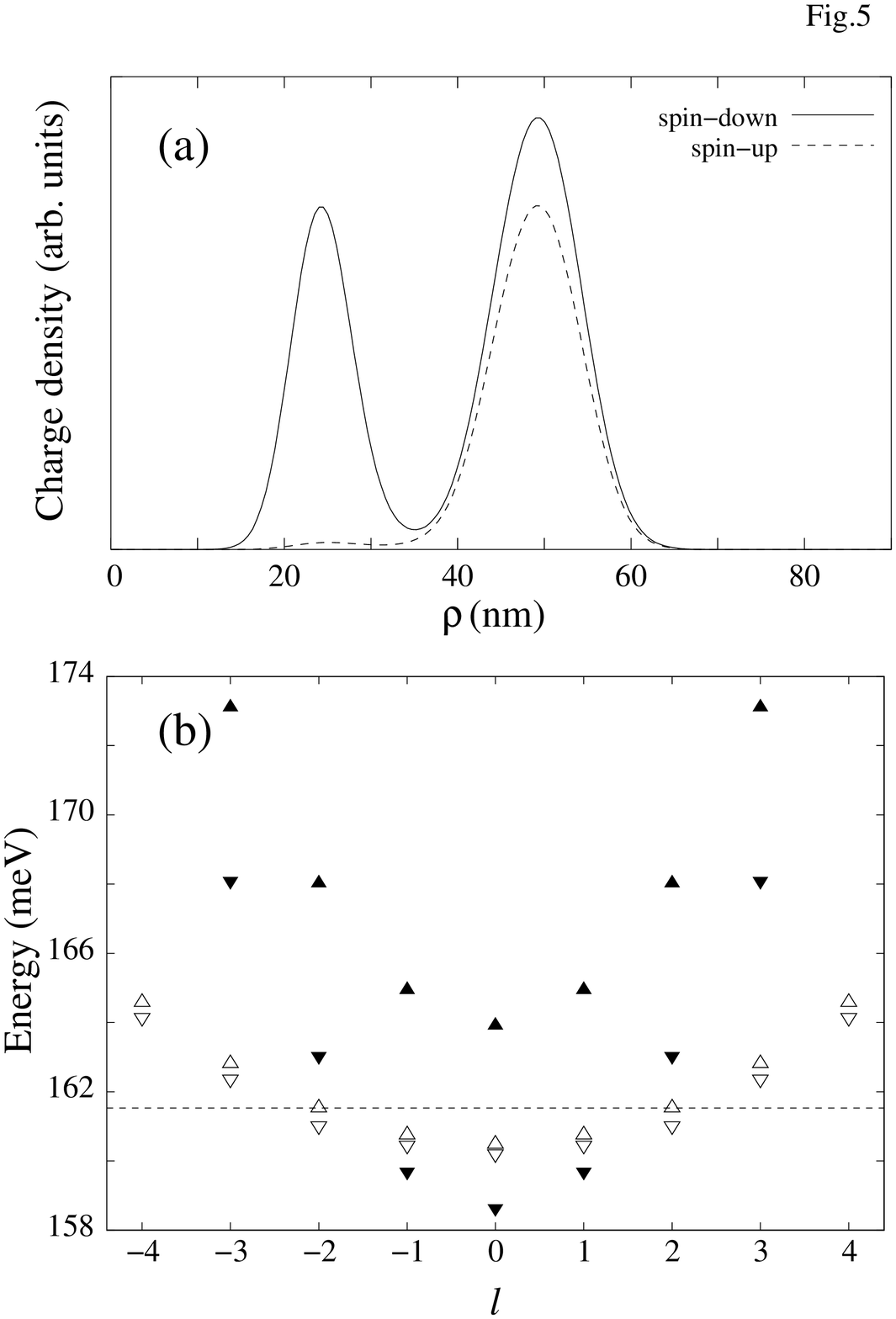} }
\caption{(a) Radial charge density distribution and (b) Kohn-Sham spin-orbital energies of the $N=12$ ground state in a 
DQR with $R_{in}=22.5$ nm and $R_{out}=47.5$ nm.
In panel (b), solid and open triangles represent orbitals localized in the inner and outer rings, respectively,
and upward(downward)-pointing triangles represent spin-up(down) orbitals. The dashed line indicates the Fermil
level.}\label{Fig5}
\end{figure}

A striking feature of the entangled DQR systems is the possibility of forming spin-dipolar ground states. 
This is illustrated in Figure \ref{Fig5}(a), where we show the
$N=12$ ground state spin-up and spin-down charge densities for the DQR with $R_{out}=47.5$ nm. 
Interestingly, the charge density in the inner ring is completely spin-polarized, while in the outer ring
it is not. To understand this phenomenon, in Figure \ref{Fig5}(b) we show the corresponding Kohn-Sham 
spin-orbitals energy diagram. Solid and open triangles represent spin-orbitals localized in the inner and outer ring, 
respectively, with upward(downward)-pointing triangles accounting for spin-up(down) orbitals.
By comparison with the independent-particle orbitals of Fig.~\ref{Fig1}, it is clear that Coulomb interaction is
now playing a major role. In particular, the energy splitting between spin-up and spin-down in the inner ring
levels is much larger than that of the outer ring. Again, this is due to the stronger Coulomb interaction in
the inner ring, which gives larger exchange-correlation
energies and thus favors the appearance of locally strongly 
spin-polarized configurations.

\begin{figure}[p]
\centerline{\includegraphics[width=8.5cm,clip]{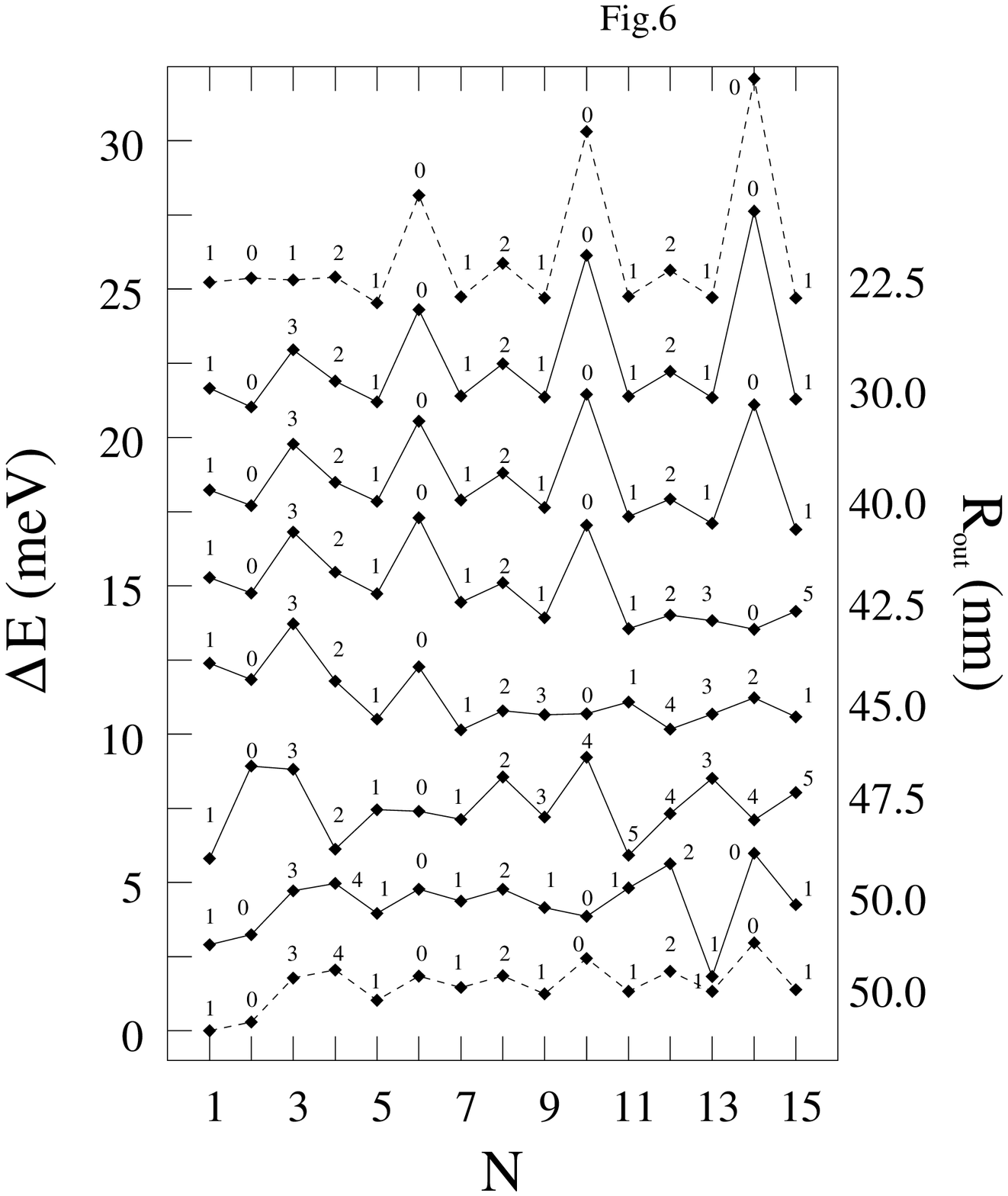} }
\caption{Addition energy spectra (full lines) versus the number of confined 
interacting electrons in DQRs with $R_{in}=22.5$ nm and changing $R_{out}$. 
Single $R = 22.5$ and $50$ nm quantum ring spectra are also shown for comparison 
(dashed lines). The spectra are offset for clarity. The number next to each peak 
indicates the spin value  $2 S_z$.}\label{Fig6}
\end{figure}

Figure \ref{Fig6} shows the addition energy spectra and spin $2\,S_z$ values of DQRs with interacting electrons.
The spectra of single quantum rings with radius $R=22.5$ and $R=50$ nm are also displayed for
comparison (dashed lines). The spectrum of a single quantum ring with $R=22.5$ nm is regular,
with maxima at closed-shell values of $N$ ($N=6,10,14$) and secondary maxima, arising from the exchange 
energy, at half-shell filling values ($N=4,8,12$).\cite{LinPRB,Emperador01}
One can realize that the $N=2$ peak, corresponding to the filling of the $n=0,l=0$ shell, is missing.
This is because the $n=0,l=0$ orbital lies very close in energy to the $n=0,l=1$ one, due to the large
radius of the quantum ring. The spin sequence is well defined by Hund's rule.
For $R_{out}=30$ nm the confinement potential is still that of a single quantum ring, but the effective mean 
radius is slightly 
increased by the exiting outer ring. As a result, irregularities are introduced around $N=3$, which now shows
a local maximum. This is due to to the formation of an exchange-favored $S_z=3/2$  three-electron
ground state, which is characteristic of quantum rings with large mean radius.\cite{ZhuPRB}
Since it could be argued that the LSDFT provides a less accurate description of
small $N$ systems, we have carried out calculations using the  configuration interaction (CI) procedure of
Ref.~\onlinecite{PlanellesEPJB} which confirm the spin-polarized ground state corresponding to $N=3$.
Up to $R_{out}=40$ nm, the spectrum remains almost constant, except for the increasing size of the $N=3$ peak
with increasing mean radius of the DQR (this is essentially a single quantum ring effect).
For $R_{out}=42.5$ and $R_{out}=45$ nm, the flattening of the addition energy spectrum for decreasing values of
$N$ reflects the localization of electron states in the outer ring. It can be observed that, once the first level of
the outer ring is occupied, the spectrum no longer displays any regular pattern, which suggests that the 
system is then ruled by Coulomb interactions.
For $R_{out} \sim 47.5$ nm, when electronic correlations play the most important role, the entire spectrum is
irregular.
Finally, for $R_{out}=50$ nm the spectrum resembles that of the single quantum ring with $R=50$ nm up to $N=9$ electrons, 
which means that the outer ring low-lying energy levels are already more stable than the inner ring ones 
(notice here that the $R=50$ nm single quantum ring spectrum is also not regular, 
i.e. it is dominated by Coulomb interactions, owing to its large radius).

\section{Conclusions}

We have investigated theoretically the electronic structure of recently synthesized self-assembled DQRs 
through the addition energy spectrum and charge density distribution. 
For most inter-ring distances, the few-electron ground states localize in either the inner or the outer ring, 
and therefore the addition energy spectrum is that of a single quantum ring. 
This is mostly a single-particle effect, due to the balance between vertical and in-plane confinement potential 
of the two rings. However, an intermediate coupling regime can be found where the energy splitting between 
single-particle states localized in either ring is smaller than Coulomb interaction energies. 
In this regime, the shell filling sequence violates Aufbau and Hund's rules, and the electron localization 
strongly depends on the number of confined particles.
Consequently, the addition energy spectrum shows an irregular structure.
Moreover, the different intensity of Coulomb interactions in the inner and outer rings may lead to ground
states with large spin-dipole moment.

\begin{acknowledgments}
Financial support from MEC-DGI Project No. CTQ2004-02315/BQU, UJI-Bancaixa Project No. P1-B2002-01, 
FIS2005-01414 (DGI, Spain), and 2005SGR00343 (Generalitat de Catalunya) are gratefully acknowledged. 
This work has been supported in part by the EU under the TMR network ``Exciting'' (J.I.C.).
\end{acknowledgments}

\clearpage

\begin{thebibliography}{20}
\bibitem{jp_dixit}
These anomalies in the orbital filling can be easily rationalized if we look at the relative energy of the Kohn-Sham 
levels vs. the number of electrons.  We do not include these figures in the paper for the sake of conciseness.
\end{thebibliography}
\end{document}